# MALWARE DETECTION TECHNIQUES FOR MOBILE DEVICES

Belal Amro, College of Information Technology, Hebron University

## ABSTRACT

*Mobile devices have become very popular nowadays, due to is portability and high performance, a mobile device became a must device for persons using information and communication technologies. In addition to hardware rapid evolution, mobile applications are also increasing in their complexity and performance to cover most the needs of their users. Both software and hardware design focused on increasing performance and the working hours of a mobile device. Different mobile operating systems are being used today with different platforms and different market shares. Like all information systems, mobile systems are prone to malware attacks. Due to the personality feature of mobile devices, malware detection is very important and is a must tool in each device to protect private data and mitigate attacks. In this paper, we will study and analyze different malware detection techniques used for mobile operating systems. We will focus on the to two competing mobile operating systems – Android and iOS. We will asset each technique summarizing its advantages and disadvantages. The aim of the work is to establish a basis for developing a mobile malware detection tool based on user profiling.*

## 1. INTRODUCTION

During the last 10 years, mobile devices technologies have grown rapidly due to the daily increase in the number of users and facilities, according to [ 1], the number of mobile users has become 4.92 billion global users in 2017. Current mobile devices can be used for many applications as camera, tablet, web browser, … etc. According to Gartner figures about smartphones, Android and iOS are the two dominant operating systems with 99.6% market share and 81.7 for Android and 17.9 for iOS [2].

A general comparison between Android and iOS mobile operating systems in provided by Aijaz sheikh et. al. [3]. Table 1 below shows some specifications of both android and iOS.

Table 1: Specifications of Android and iOS

| Specification | Android | iOS |
|---|---|---|
| Developer | Google | Apple |
| OS family | Linux | OS X, Unix |
| Initial release | 23/9/2008 | 29/7/2007 |
| Programming language | C, C++, Java | C, C++, objective-C |
| Source model | Open source | Closed source |
| Latest stable release | Oreo 8 , august 21, 2017 | 11 , September , 19, 2017 |

Android operating system is divided into four layers as shown in Figure 1, the Linux kernel is the bottom layer responsible for abstraction of device hardware. The libraries layer contains a set of libraries including WebKit, Libc, and SSl. Android libraries includes Java-based libraries such as android view and android widget. Application framework layer provides higher level services to applications in terms of Java classes. The top layer is called application layer where applications are written to be installed.

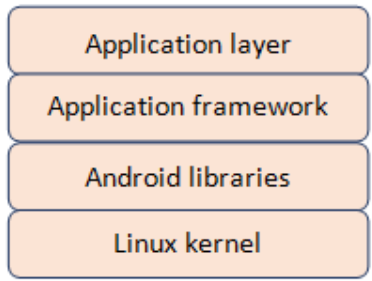

Figure1: Android architecture

The iOS architecture is shown in Figure 2. The Cocoa Touch layer contains frameworks for iOS apps. Media layer contains the graphics, video, and audio technologies for iOS apps. The core services layer contains the fundamental system services for iOS apps. At bottom, the core OS layer contains the low-level features that most other technologies are built upon [4]

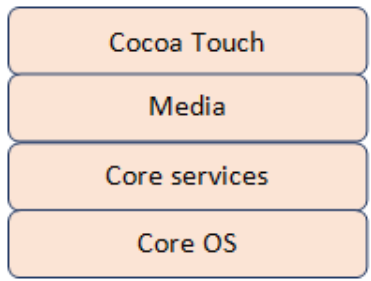

Figure 2: iOS architecture

In terms of application distribution, Android applications are mostly distributed through google play where more than half of the applications are free. Apple applications are distributed through App store, almost quarter of the applications are for free. An important issue is that all iOS applications at App store are scrutinized before they are released. The later step made App store applications more reliable than those at google play [5].

The rest of the paper is organized as follows, a summary of mobile malware is provided in Section 2. Section 3 describes malware spreading techniques. Malware evasion techniques are provided in Section 4. The detection techniques used by antimalware programs are describes in Section 5. At last, Section 6 summarizes the work done in this paper.

**MOBILE MALWARE ANALYSIS:**

In this section, we provide a summary of mobile malwares including Trojans, Back doors, Ransomwares, Botnets, and Spyware. Besides, a statistical data about malwares and their distribution is provided as well.

## MOBILE MALWARES:

As reported by Skycure [31], one third of mobile devices has a medium to high risk of data disclosure, Android devices are nearly twice likely to have a malware compared to iOS devices. in this subsection, we will explain some of the most important mobile malwares.

## TROJANS:

Trojan is a software that appears to the user to be benign application however, it performs malicious acts in the back ground[6]. Trojan are used to help attacking a system by performing acts that might compromise security of the system and hence enables hacking it easily. Examples of Trojans are FakeNetflix [7], which collects users credentials for Netflix account in Android environments. KeyRaider is a Trojan that was used to steal Apple IDs and passwords[17].

## BACK DOORS – ROOT EXPLOITS

Backdoors exploits root privileges to hide a malware from antiviruses. Rage against the cage (RATC) is one of the most popular Android root exploits which gain full- control of device [8]. If the root exploit gains root privilege, the malware become able to perform any operation on the device even the installation of applications keeping the user unaware of this act [9]. In iOS, Xagent is a Trojan that opens a back door and steals information from the compromised device [16]

## RANSOMWARE

Ransomware prevents the users from accessing their data by locking the device or encrypting the data files, until ransom amount is paid. FakeDefender.B [10] is a malware pretending to be Avast antivirus. It locks the victim's device for the sake of money. An iOS ransomware was reported in 2017, scammers exploited Safari bug used for pop-up [35].

## BOTNETS

A "bot" is a type of malware that enables an attacker to take control over an affected Mobile device, it is also known as "Web robots", they are part of a network of infected machines, known as a "botnet", which is typically made up of all victim mobile devices across the globe. Geinimi [11] is one of the Android botnets.

## SPYWARE

A spyware is simply a spying software. It runs unnoticed in the background while it collects information, or gives remote access to its author. Nickspy [12] and GPSSpy [13] are examples of Android spyware that monitors the user's confidential information and sends them to the owner. An example of an iOS Spyware is Passrobber[16] , which is capable of intercepting outgoing SSL communications, it then checks for Apple IDs and passwords, and can send these stolen credentials to a C&C sever.

## MOBILE MALWARE STATISTICS:

In this section, we provide some statistics about mobile malware attacks. The number of mobile malwares is increasing dramatically last two years. According to MacAfee LABs [28], the number of malwares exceeded 16,000,000 in first quarter of 2017 as shown in Figure 3.

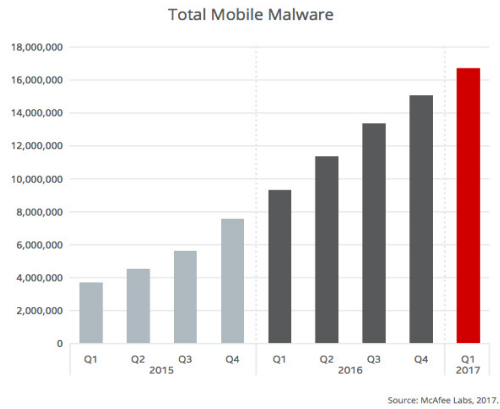

Figure 3: Total mobile malware

By looking at the global mobile malware infection rate reported by MacAfee LABs 2017, Figure 4 shows a significant increase in the infection rate for the first quarter of the year 2017.

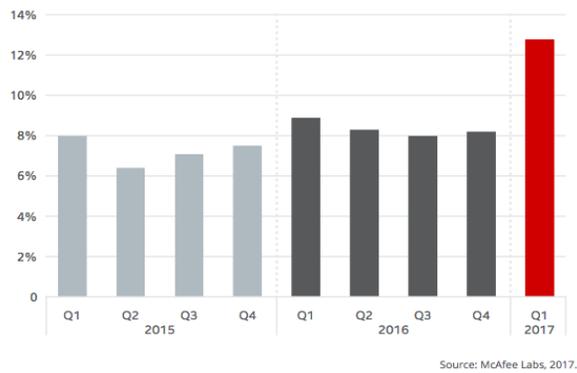

Figure 4: global mobile malware infection rates

Kaspersky Labs [32] reported the distribution of new mobile malware in the years 2015 and 2016 as shown in Figure 5:

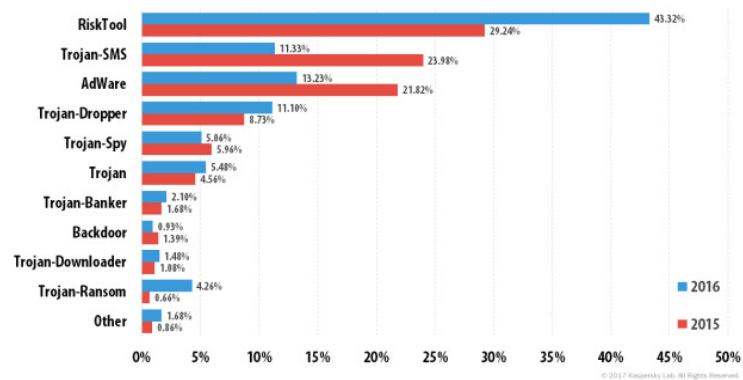

Figure 5: distribution of mobile malware

As reported by LookingGlass [33], "in 2015, the threat actors shift their tactics to smaller targets with mobile-ransomware focusing more on individuals and less on corporations. The bring your own Device (BYOD) environment became more pervasive with organizations realizing the importance of establishing concrete BYOD policies".

A survey conducted by Dimensional research [34] on security professional reported that security professionals are unprepared and not confident about arising security issues, it also reported that mobile devices are to come under increasing attacks.From this section, we realize that mobile threats are increasing rapidly and are more focused on targets. This made us to predict a huge damage in the near future unless efficient tools are developed and used.

## MALWARE SPREADING TECHNIQUES

To mitigate malware attacks, we should be aware of malware spreading techniques. In this section, we categorize malware spreading techniques including repackaging, drive by download, dynamic payloads, and stealth techniques.

### REPACKAGING

Malware authors repackage popular mobile applications in official market, and distribute them on other less monitored third party markets. Repackaging includes the disassembling of the popular benign apps, then appending the malicious content and finally reassembling. This is done by reverse-engineering tools. TrendMicro report have shown that 77% of the top 50 free apps available in Google Play are repackaged [14].

### DRIVE BY DOWNLOAD

Drive by Download refers to an unintentional download of malware in the background. It Occurs when a user visits a website that contains malicious content and downloads malware into the device. Android/NotCompatible [15] is the most popular mobile malware of this category.

### DYNAMIC PAYLOADS

Uses dynamic payload to download an embedded encrypted source in an application. After installation, the application decrypts the encrypted malicious payload and executes the malicious code [16].

### STEALTH MALWARE TECHNIQUES

Stealth Malware Technique refers to an exploit of hardware vulnerabilities to obfuscate the malicious code to easily bypass the anti-malware. Different stealth techniques such as key permutation, dynamic loading, native code execution, code encryption, and java reflection are used to attack the victim's device[16].

### MALWARE EVASION TECHNIQUES

Kaspersky LABs reported in their 2016 year findings [1] that malware creators have used new ways to bypass Android protection mechanisms. Malware creators need to constantly monitor mobile security techniques and develop new techniques to avoid detection. These techniques are called evasion techniques and are listed below [29]:

**Anti-security techniques:** these techniques are used to avoid detection by security devices and programs as anti-malwares, firewalls, and any other tools that protect the environment.

**Anti-sandbox techniques:** sandboxing is a technique used to separate running programs and hence to avoid any harm from unverified programs to the computer system. Anti-sandbox technique is used to detect automatic analysis and to avoid report on the behavior of malware. This can be done by detecting registry keys, files, or processes related to virtual environments.

**Anti-analyst techniques**: in these techniques, a monitoring tool is used to avoid reverse engineering. The tools might be process explorer or Wireshark to perform monitoring and to detect malware analyst.

Malware creators might use two or three of the above techniques to make detection more difficult. Figure 6 shows the popularity of evasion techniques used by malware creators:

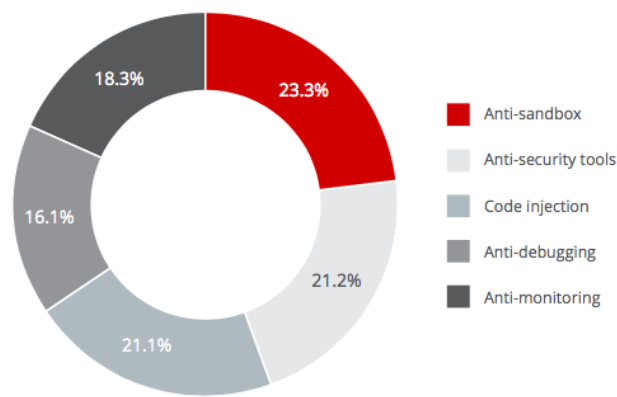

Figure 6: Evasion techniques used by malwares

## MALWARE DETECTION TECHNIQUES:

In this section, we analyze the state-of-the-art malware detection techniques for mobile phones. We categorized them in two categories according to the basis they rely on when detecting for malwares. The categories are statics and dynamic techniques

## STATIC TECHNIQUES:

Static techniques rely on the source code of an application to classify it accordingly without having the application being executed. These techniques are classified into one of the following classes according to the basis they rely on for analyzing the source code:

## SIGNATURE BASED APPROACH

This method extracts the semantic patterns and creates a unique signature [18]. A program is classified as a malware if its signature matches with existing signatures. It is a very fast method for detecting malware, however, it can be easily circumvented by code obfuscation.  IT can only identify the existing malwares and fails against the unseen variants of malwares. It also needs immediate update of malware signatures.

**PERMISSION BASED ANALYSIS:**

Permissions requested by the application plays a vital role in governing the access rights. By default, apps have no permission to access the user's data and effect the system security. User must allow the app to access all the required resources during installation process. It is worth mentioning that developers must mention the permissions requested for the resources. But not all declared permissions are necessarily required permissions as shown in [19].

Permission based detection is fast in application scanning and identifying malware but do not analyze other files which contain the malicious code. Also a very small difference in permissions exists between malicious and benign applications, hence, permission based methods require second pass to provide efficient malware detection.

**VIRTUAL MACHINE ANALYSIS:**

In mobile application, a virtual machine is used to test the byte code of a particular application. Bytecode analysis tests the app behavior and analyses control and data flow which might be helpful in detecting dangerous functionalities performed by malicious applications. Plenty of virtual machine application have been implemented for mobile devices, specially for android systems. DroidAPIMiner [20], identifies the malware by tracking the sensitive API calls.

Limitations of virtual machine analysis is that analysis is performed at instruction level and consumes more power and storage space.

**DYNAMIC TECHNIQUES:**

In dynamic analysis, an application is examined during execution and then classified according to one of the following techniques. The classification is done according to the behavior of the detection mechanism.

**ANOMALY BASED**

Anomaly based analysis is based on watching the behavior of the device by keeping track of different parameters and the status of the components of the device. Andromly is a behavior based malware detection technique [21]. To detect a malware, Andromly continuously monitors the different features of the device state such as battery level, CPU usage, network traffic, etc. Measurements are taken during running and are then supplied to an algorithm that classifies them accordingly. CrowDroid [22] and AntiMalDroid [23], are two different anomalies based tools used for malware detection in Android devices. The first depends on analyzing system calls' logs while the latter analyzes the behavior of an application and then generates signatures for malware behavior. SMS Profiler and iDMA are two tools used to detect illegitimate use of system services in iOS[24].

**TAINT ANALYSIS**

Taintdroid [25] is a tool that tracks multiple sources of sensitive data and identifies the data leakage in mobile applications. The tool labels sensitive data and follows the data moving from the device. Taintdroid provides efficient tracking of sensitive data, unfortunately, it does not perform control flow tracking.

International Journal of Mobile Network Communications & Telematics ( IJMNCT) Vol.7, No.4/5/6, December 2017

**EMULATION BASED**

DroidScope [26] is an emulation based tool used to dynamically analyze applications based on Virtual Machine Introspection. It monitors the whole system by being out of execution environment, hence malwares will not be able to detect existence of anti-malware installed on the device.

Another emulation based tool provided by Blaising et al. [27] and called Android Application Sandbox (AASandbox). AASandbox detects the malicious applications by using static and dynamic analysis. The effect of the tool is limited to sandbox for security reasons. The tool dynamically analyzes the user behavior such as touches, clicks and gestures etc. Unfortunately, the tool cannot detect new malwares.

## 2. SUMMARY

Malware attacks have been growing rapidly last 10 years, these attacks targeted all technology device including mobile phones. Due to the personality of the mobile usage and the sensitive data they might contain, safeguards against malwares must be implemented. In this paper, we introduced different types of attacks on the top two competing mobile operating systems – Android and iOS. We also introduced the techniques used to deliver mobile malwares, and provided up-to-date statistics for malware attacks in the last 3 years. We then introduced the most common malware detection techniques used for mobile applications. We also pinpointed and discussed the weakness in each malware detection technique. We will be working on developing a new malware detection tool for mobile devices that can be used efficiently based on mobile user profiling.